\documentclass[cameraready]{Interspeech}

\title{Audio2Tool: Speak, Call, Act - A Dataset for Benchmarking Speech Tool Use}

\author[affiliation={1}, orcid=0009-0001-8720-4406, equalcontribution, correspondingauthor]{Ramit}{Pahwa}
\author[affiliation={1}, orcid=0000-0002-5874-1800, equalcontribution]{Apoorva}{Beedu}
\author[affiliation={1}]{Parivesh}{Priye}
\author[affiliation={1}]{Rutu}{Gandhi$^\dagger$}
\author[affiliation={1}]{Saloni}{Takawale$^\dagger$}
\author[affiliation={1}]{Aruna}{Baijal}
\author[affiliation={1}]{Zengli}{Yang}

\address{
    $^1$ Rivian and Volkswagen Group Technologies 
}

\email{\{ramitpahwa, apoorvabeedu, pariveshpriye, rgandhi, salonitakawale, abaijal, zengliyang\}@rivianvw.tech}

\keywords{speech recognition, spoken language understanding, function/tool calling}

\usepackage{comment}
\usepackage{algorithm}
\usepackage{algpseudocode}
\usepackage{float}
\usepackage{amsmath}
\usepackage{booktabs}
\usepackage{multirow}
\usepackage{tabularx}
\usepackage{makecell}

\usepackage{array}
\newcolumntype{P}[1]{>{\centering\arraybackslash}p{#1}}
\newcolumntype{L}[1]{>{\raggedright\arraybackslash}p{#1}}
\newcolumntype{C}[1]{>{\centering\arraybackslash}p{#1}}

\usepackage[capitalise]{cleveref}
\usepackage{subcaption}

\begin{document}
\maketitle

\bgroup
\renewcommand{\thefootnote}{}
\footnotetext{$^\dagger$These authors contributed equally.}
\footnotetext{Submitted for review to Interspeech 2026.}
\egroup

\begin{abstract}
    Voice assistants increasingly rely on SpeechLMs to interpret spoken queries and execute complex tasks, yet, existing benchmarks lack domain breadth, acoustic diversity, and compositional reasoning complexity to evaluate tool-calling performance.
We introduce Audio2Tool, a large scale dataset comprising ~30,000 queries designed to assess tool-calling capabilities of SpeechLMs across three primary domains: Smart Car, Smart Home, and Wearables.
Our benchmark features a multi-tier complexity hierarchy, ranging from simple direct commands to complex multi-intent and needle-in-a-haystack extraction to isolate distinct failure modes.
To ensure realism, we employ zero-shot voice cloning TTS and diverse noise profiles to simulate in-the-wild conditions.
Evaluations of state-of-the-art SpeechLMs and ASR–LLM pipelines show strong performance on simple commands but significant degradation under compositional and acoustic challenges.
Code and dataset are publicly available on the project page: \href{https://audio2tool.github.io/}{https://audio2tool.github.io/}.
\end{abstract}

\section{Introduction}
\label{sec:intro}

Recent advances in Speech Large Language Models (SpeechLMs) are transforming voice assistants from simple intent recognition systems into end-to-end, audio-native agents capable of directly invoking tools from raw speech.
In this setting, the model performs \textit{audio-native function calling}, mapping raw acoustic signals directly to executable API calls.
Prior approaches relied on cascaded pipelines (ASR-LLM), where an Automatic Speech Recognition (ASR) system produced text that was subsequently processed by a Natural Language Understanding (NLU) module.
These pipelines are inherently limited: ASR errors propagate into tool retrieval, and the distillation of speech into text discards paralinguistic features like prosody and intonation that are essential for intent disambiguation in noisy environments.

Despite the growing capabilities of SpeechLMs, the field still lacks a large scale, realistic benchmarks that systematically evaluates tool-calling performance and robustness under diverse acoustic conditions.
Recent advances in tool calling have driven the development of multiple datasets and benchmarks \cite{chen2024voicebench,zhong2025complexfuncbench,patil2025berkeley}.
While text-based tool calling has been extensively benchmarked through datasets like Berkeley Function Calling Leaderboard (BFCL) \cite{patil2025berkeley}, the transition to the audio domain remains a significant bottleneck for wide adoption in real-world applications.
In speech-driven interfaces, a model must not only parse semantic intent but also navigate the inherent phonetic ambiguities and acoustic variabilities of spoken language.
While current speech-centric benchmarks such as AudioBench \cite{wang2025audiobench}, VoiceAgentBench \cite{jain2025voiceagentbench} etc., have laid the groundwork for evaluating speech-language models (SLMs), these datasets suffer from limited taxonomic breadth and a lack of acoustic diversity. 
Many are restricted to high-resource languages or curated laboratory conditions, failing to reflect the ``in-the-wild", naturalistic environments where voice assistants actually operate. 
Furthermore, they typically lack a structured, multi-tier query hierarchy, which is essential for evaluating models' ability to reason across varying levels of granularity: from simple API triggers to complex, interdependent function chains.
There is a critical need for a large-scale, acoustically diverse dataset that challenges models to bridge the gap between raw audio signals and sophisticated tool orchestration.

In this work, we introduce \textbf{Audio2Tool}, a dataset comprising approximately 30,000 queries designed to push the boundaries of audio-based tool calling. 
Our dataset is distinguished by its large scale and a multi-tier query architecture that covers an expansive taxonomy of real-world use cases. 
Unlike previous efforts, we prioritize realistic deployment scenarios by integrating extensive variability in speaker prosody and environmental conditions. 
Our contributions include:

\begin{itemize}
    \item Scale and Taxonomy: A dataset of approx. 30,000 queries across a broad functional hierarchy designed specifically tools for smart car, smart home, wearables.
    \item Multi-tier Reasoning: Queries designed to test both single-turn and complex multi-step tool execution.
    \item Acoustic Realism: Inclusion of diverse noise, speaker prosody, and environmental perturbations.
    \item We provide extensive baseline experiments that establish new standards for robustness in the audio-to-tool pipeline. 
\end{itemize}

By providing a rigorous testing ground for these capabilities, we aim to accelerate the development of voice agents that are not only intelligent in their reasoning but also resilient in their perception. 
To best of our knowledge this is the first benchmark which is designed with realistic tool call and acoustic conditions representative of the category. 

\begin{figure*}[!t]
\centering
\includegraphics[width=1\linewidth]{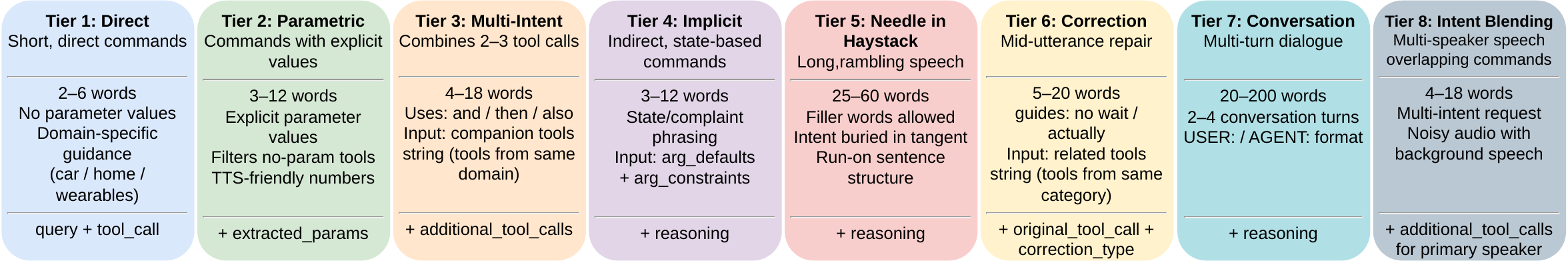}
\caption{Overview of the Audio2Tool query generation pipeline detailed for different tiers.
}
\vspace{-1em}
\label{fig:tier_creation}
\end{figure*}
\section{Background}
\label{sec:background}


With the growing interest in developing
systems capable of end-to-end interaction, Spoken Language Understanding (SLU) has expanded to also encompass 
the ability to invoke executable function calls directly from speech.
Going beyond traditional  cascaded pipelines, recent SpeechLMs enable audio native reasoning, allowing models to directly select relevant tools\cite{patil2025berkeley,wang2025audiobench}. 

Early benchmarks such as the Air Travel Information System corpus \cite{hemphill1990atis} focused on flight-related queries, while later datasets such as the Spoken Language Understanding Resource Package (SLURP) \cite{bastianelli2020slurp}, the Spoken Task-Oriented Semantic Parsing dataset (STOP) \cite{stop2021}, and the Multi-Intent Automotive Cabin Spoken Language Understanding dataset (MAC-SLU) \cite{peng2025mac} covered multi-domain, compositional, and multi-intent settings.
While these benchmarks are challenging, they do not directly evaluate executable tool invocation from speech.\\
\textbf{From SLU semantics to executable tool calling:} 
Traditional SLU systems produce intermediate semantic representations such as intents, slots, or structured parses.
Tool calling extends this to \emph{executable} structure: i.e., selecting an API and populating valid arguments from speech. 
This introduces stricter correctness constraints than standard SLU, including schema validity, high-fidelity entities (e.g., identifiers and timestamps), and ordered multi-step dependencies, such that semantically plausible outputs may still fail if any component is invalid. 
As such, benchmarks such as the Berkeley Function Calling Leaderboard (BCFL) \cite{patil2025berkeley}, BFCL Audio \cite{bfcl_audio}, VoiceAgentBench \cite{jain2025voiceagentbench}, and MFCL \cite{mao2025mfcl} evaluate function calling across text, audio, and multimodal settings, but provide limited support for analysis of fine-grained audio-to-tool failure modes.

In contrast, Audio2Tool is designed for domain-grounded and diagnostic evaluation of \emph{audio-to-tool} capabilities.
Complementary to the aforementioned function-calling leaderboards \cite{patil2025berkeley} and audio/multi-modal tool-use benchmarks \cite{bfcl_audio,mao2025mfcl}, it organizes tools and APIs into taxonomies across three application domains and introduces a eight-tier query curriculum that progressively assesses tool selection, argument grounding, and multi-step composition.
Finally, we generate speech
with controlled perturbations to assess robustness and failure modes that are specific to speech-driven tool invocation.
\vspace{-0.5em}
\section{Benchmark}
\label{sec:method}

Agentic tool-oriented interactions require capabilities such as audio-native function calling and mapping speech directly to executable API calls. 
However, evaluating such capabilities requires benchmarks to capture the interplay of speech prosody, acoustic variability, contextual cues, and multi-domain API constraints.
To address this need, we introduce a benchmark and an evaluation framework designed to assess audio-to-tool systems beyond intent-level accuracy.
We accomplish this by incorporating a hierarchical tool taxonomy, multi-tiered query complexity, and carefully curated, and manually verified synthetic audio generation strategies.

\subsection{Tool Taxonomy}
To enable structured evaluation of tool-calling behavior across domains, we construct a unified tool taxonomy grounded in real-world APIs, including standard android automotive functionalities, smart home devices, and wearable devices.
We curate 152 verified functions spanning three domains: Smart Car, Smart Home, and Wearables. 
These functions are organized into 23 categories designed to reflect the operational subsystems of modern IoT systems.
The \textbf{Smart Car} domain comprises 14 categories, spanning climate control, driving dynamics, charging, and vehicle maintenance. 
Meanwhile, the \textbf{Smart Home} domain covers lighting, appliances, automation sensors, and security-related functions, and \textbf{Wearables} focuses on health tracking and device connectivity.

Our taxonomy design is guided by two requirements: 
\textit{(i)} {Operational intent}: distinguishing state-altering commands (e.g., \emph{Set temperature}) from passive monitoring actions (e.g., \emph{Check battery}); 
and
\textit{(ii) } {Domain specificity:} ensuring specialized categories capture device-specific behaviors, such as driving dynamics in vehicles or activity tracking in wearables. 


\begin{figure*}[!t]
    \centering
    \begin{subfigure}[t]{0.45\linewidth}
        \centering
        \includegraphics[width=\linewidth,height=3cm]{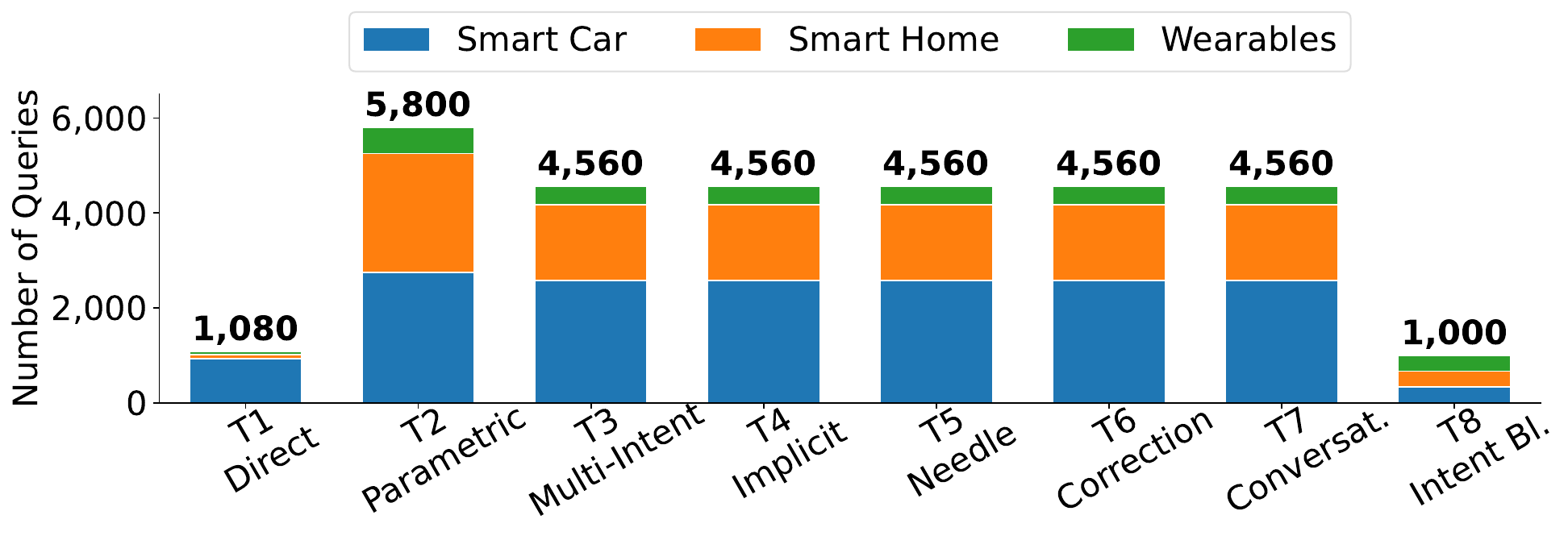}
        \caption{Number of queries per complexity tier.}
        \label{fig:queries_per_tier}
    \end{subfigure}
    \hfill
    \begin{subfigure}[t]{0.24\linewidth}
        \centering
        \includegraphics[width=\linewidth,height=3cm,keepaspectratio]{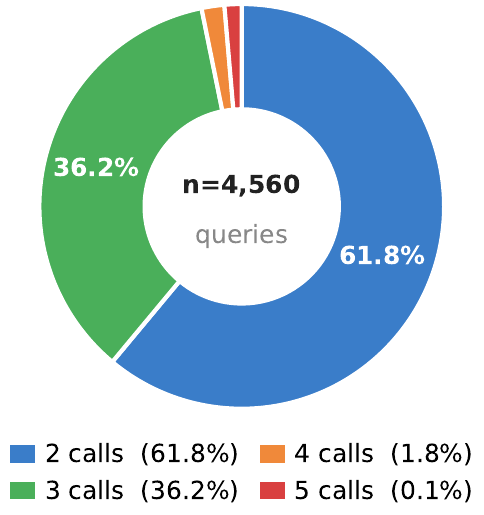}
        \caption{Intent count in Tier~3.}
        \label{fig:tier3_call_count}
    \end{subfigure}
    \hfill
    \begin{subfigure}[t]{0.3\linewidth}
        \centering
        \includegraphics[width=\linewidth,height=3cm,keepaspectratio]{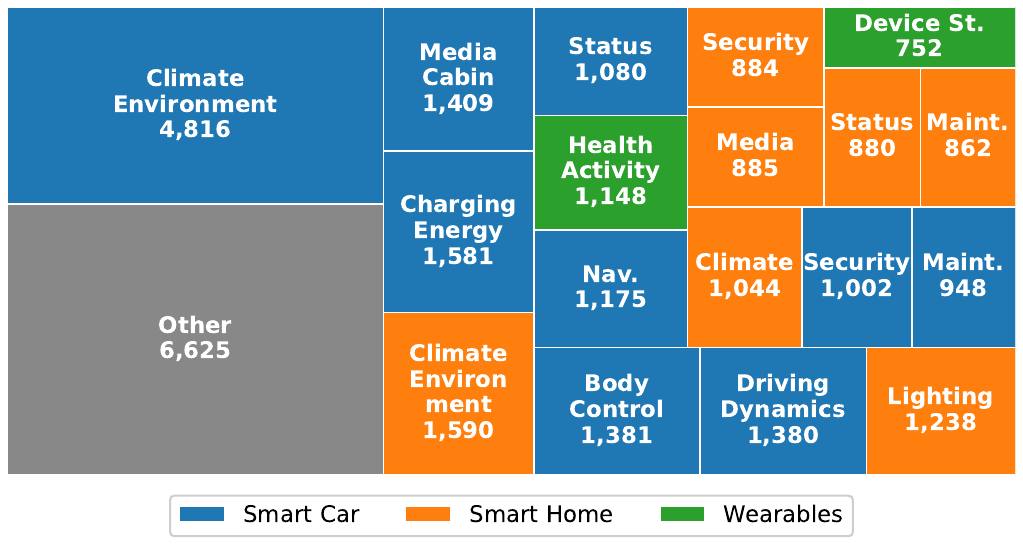}
        \caption{Category distribution.}
        \label{fig:category_distribution}
    \end{subfigure}
    \caption{Query and domain distributions across complexity tiers.}
    \vspace{-1em}
    \label{fig:queries_and_domains}
\end{figure*}

\begin{table}[!t]
\centering
\caption{Comparison of SLU datasets.}
\label{tab:slu_comparison}
\resizebox{0.8\columnwidth}{!}{
\begin{tabular}{L{.25\linewidth} C{.15\linewidth} C{0.15\linewidth} C{0.15\linewidth} C{0.25\linewidth}}
\hline
\textbf{Dataset} & \textbf{Domain} & \textbf{Intent} & \textbf{Slot} & \textbf{Multi-intent} \\
\midrule
ATIS\cite{hemphill1990atis} & 1 & 16 & 41 & \texttimes \\
SNIPS\cite{coucke2018snips} & 2 & 7 & 4 & \texttimes \\
FSC [3] & 2 & 6 & 2 & \texttimes \\
SLURP\cite{bastianelli2020slurp} & 18 & 46 & 56 & \texttimes \\
MixATIS\cite{nguyen2023joint} & 1 & 16 & 41 & \checkmark \\
MixSNIPS\cite{qin2020agif} & 2 & 7 & 4 & \checkmark \\
MAC-SLU\cite{peng2025mac} & 8 & 81 & 192 & \checkmark \\
\midrule
\textbf{Audio2Tool} & \textbf{3} & \textbf{152} & \textbf{22} & \checkmark \\
\bottomrule
\end{tabular}
}
\vspace{-2em}
\end{table}

\textbf{Query Generation:}
Queries are constructed by utilizing tools from the taxonomy and varying query length and complexity.
To this end, we define eight complexity tiers that progressively increase both linguistic complexity and functional reasoning for tool calling.
\Cref{fig:tier_creation} provides an overview for the different tier query generation.
\begin{itemize}
\item \textbf{Tier 1 (Direct):} Short, direct commands (2–6 words; no fillers); e.g., ``Open the trunk.''
\item \textbf{Tier 2 (Parametric):} Slot-filling commands with constrained parameters; e.g., ``Set the temperature to 72.''
\item \textbf{Tier 3 (Multi-Intent):} Single utterances expressing multiple intents; e.g., ``Defrost the windshield and find a charger.''
\item \textbf{Tier 4 (Implicit Reasoning):} Contextual utterances requiring pragmatic inference; e.g., ``I’m freezing.''
\item \textbf{Tier 5 (Needle-in-a-Haystack):} Extraction of tool calls from long-form audio containing substantial irrelevant context.
\item \textbf{Tier 6 (Correction):} Handling mid-utterance revisions and state corrections; e.g., ``Set an alarm for 7\ldots wait, make it 8.''
\item \textbf{Tier 7 (Conversation):} Multi-turn dialogue (4–8 turns) requiring persistent intent tracking and state maintenance.
\item \textbf{Tier 8 (Intent Blending):} Disambiguating primary user commands from background speech containing secondary valid intents (e.g., "Navigate home" (primary speaker), "Set temperature to 72" (secondary speaker).
\end{itemize}

Overall, these tiers are designed to cover a wide gamut of interactions, and to isolate distinct failure modes that occur in real-world deployments.  
Tiers~1--2 evaluate basic tool-calling capabilities, including intent recognition and correct extraction of parameters from explicit commands. 
Meanwhile, Tiers~3--4 test the model’s ability to handle multiple tool calls in a single utterance or to infer the intended action when it is only implied. 
Tiers~5--8 focus on robustness under more realistic conditions, including irrelevant context, mid-utterance corrections, and multi-turn dialogue requiring consistent intent tracking. 
Unlike Tier 5 (which filters irrelevant noise), Tier 8 requires the model to distinguish between the primary user's voice and a ``distractor" intent (e.g., a podcast, radio, or another person in the room) to avoid accidental execution of the wrong command. 
The key challenge is speaker diarization and context-aware filtering, to ensure the system only responds to the primary actor.

The distribution of queries across complexity tiers and tool domains is illustrated in \Cref{fig:queries_per_tier,fig:category_distribution}, with \Cref{fig:tier3_call_count} further breaking down the intent composition within Tier~3. 
Overall, the dataset has a balanced distribution across Tiers 3–7, each containing 4,560 queries, while simpler direct requests (Tier 1) and highly parameterized queries (Tier 2) account for 1,080 and 5,800 instances, respectively. 
This design ensures broad coverage of interaction complexity, from direct commands to multi-turn conversational scenarios.
All queries are generated using closed-source LLMs, including GPT-5.2, Gemini 2.5 Pro, and Claude Opus.
We evaluate these queries for their accuracy, difficulty, and variability, using an LLM-as-the-judge setup comprising GPT-5.1 and Gemini-2.5-pro. 
For queries where at least one LLM failed, the query and the ground truth tools were manually checked for correctness.

Across tiers, the Smart Car domain consistently constitutes the majority of the queries. 
This distribution reflects our focus on high-stakes, hands-free environments where accurate tool orchestration is critical. 
By prioritizing the automotive domain, we ensure the benchmark evaluates the model's ability to handle the dense parameter requirements and specialized subsystems (e.g., climate control, vehicle maintenance).

As further shown in \Cref{tab:slu_comparison}, compared to prior SLU benchmarks, Audio2Tool spans fewer domains but supports a substantially larger intent space and explicitly models multi-intent interactions. 
We emphasize functional diversity and compositional reasoning, aligning the dataset with the practical requirements of modern tool-calling systems rather than traditional slot-filling paradigms.
\begin{table}[!t]
\centering
\caption{Datasets for speaker diversity and audio conditions.}
\label{tab:voice_datasets}
\setlength{\tabcolsep}{3pt}
\renewcommand{\arraystretch}{1.3}
\resizebox{0.8\linewidth}{!}{%
\begin{tabular}{L{.35\linewidth} L{.175\linewidth} L{0.3\linewidth} L{0.175\linewidth}}
\hline
\textbf{Dataset} & \textbf{Speakers} & \textbf{Regions} & \textbf{Selected Speakers} \\
\hline
SPGISpeech 2.0 \cite{grossman2025spgispeech}& 41,593 & US, Asia, LatAM & 100 \\
Emilia-Yodas \cite{yodasspeakerpool2026} & 7,092 & US, China & 100 \\
3D Speaker \cite{zheng20233d} & 10,000 & China & 30 \\
VoxPopuli \cite{wang2021voxpopuli} & 1,180 & US, Europe & 100 \\
\hline
\end{tabular}%
}
\vspace{-1em}
\end{table}

\section{TTS Voice Generation and Processing}
\begin{table*}[t]
\centering
    \caption{
    Tool-calling performance across complexity tiers. Acc, EM, and F1 denote tool accuracy, exact match, and  slot/parameter F1 scores respectively. 
    F1 is undefined for Tier~1 queries, as no parameters exist.
    }
    \label{tab:tool_calling_results}
    \setlength{\tabcolsep}{3pt}
    \renewcommand{\arraystretch}{1.3}
    \resizebox{1\textwidth}{!}{%
    \begin{tabular}{
    L{0.2\textwidth} |
    C{0.0333\textwidth} C{0.0333\textwidth} C{0.0333\textwidth} |
    C{0.0333\textwidth} C{0.0333\textwidth} C{0.0333\textwidth} |
    C{0.0333\textwidth} C{0.0333\textwidth} C{0.0333\textwidth} |
    C{0.0333\textwidth} C{0.0333\textwidth} C{0.0333\textwidth} |
    C{0.0333\textwidth} C{0.0333\textwidth} C{0.0333\textwidth} |
    C{0.0333\textwidth} C{0.0333\textwidth} C{0.0333\textwidth} |
    C{0.0333\textwidth} C{0.0333\textwidth} C{0.0333\textwidth} |
    C{0.0333\textwidth} C{0.0333\textwidth} C{0.0333\textwidth}
    }
    \hline
    \multirow{2}{*}{\textbf{Model}} &
    \multicolumn{3}{c|}{\textbf{Tier-1}} &
    \multicolumn{3}{c|}{\textbf{Tier-2}} &
    \multicolumn{3}{c|}{\textbf{Tier-3}} &
    \multicolumn{3}{c|}{\textbf{Tier-4}} &
    \multicolumn{3}{c|}{\textbf{Tier-5}} &
    \multicolumn{3}{c|}{\textbf{Tier-6}} &
    \multicolumn{3}{c|}{\textbf{Tier-7}} &
    \multicolumn{3}{c}{\textbf{Tier-8}} \\
    \cline{2-25}
    & \textbf{Acc} & \textbf{EM} & \textbf{F1}
    & \textbf{Acc} & \textbf{EM} & \textbf{F1}
    & \textbf{Acc} & \textbf{EM} & \textbf{F1}
    & \textbf{Acc} & \textbf{EM} & \textbf{F1}
    & \textbf{Acc} & \textbf{EM} & \textbf{F1}
    & \textbf{Acc} & \textbf{EM} & \textbf{F1}
    & \textbf{Acc} & \textbf{EM} & \textbf{F1}
    & \textbf{Acc} & \textbf{EM} & \textbf{F1} \\
    \toprule
    
    Qwen 1.7B
    & 74.6 & 74.6 & --
    & 35.7&  3.2 & 5.0
    & 48.2 & 30.4 & 36.7
    & 18.2 & 8.9  & 14.9
    & 68.4 & 35.6 & 41.9
    & 62.3 & 36.8 & 44.5
    & 43.1 & 20.2 & 29.8
    & - & - & - \\
    
    Qwen 4B
    & 80.2 & 80.2 & --
    &41.6 & 6.7 & 13.1
    & 54.6 & 34.1 & 39.2
    & 21.7 & 10.1 & 16.2
    & 73.9 & 39.8 & 46.2
    & 69.1 & 41.5 & 49.1
    & 48.9 & 24.3 & 33.1
    & - & - & - \\
    
    Qwen 8B
    & 85.6 & 85.6 & --
    & 77.1 & 10.1 & 19.3
    & 61.3 & 38.7 & 43.6
    & 26.8 & 12.6 & 18.9
    & 80.5 & 44.9 & 51.3
    & 75.2 & 46.8 & 54.6
    & 55.4 & 28.7 & 37.9
    & - & - & - \\
    
    Gemma 12B
    & 84.1 & 84.1 & --
    & 78.3 & 12.3 & 20.4
    & 59.2 & 37.6 & 42.7
    & 27.4 & 14.5   & 19.1
    & 80.8 & 45   & 51.5
    & 73.6 & 45.9 & 53.8
    & 55.7 & 29.9 & 38.8
    & - & - & - \\
    
    \midrule
    
    whisperv3 + Qwen 1.7B
    & 68.2 & 68.2 & --
    & 33.6 & 2.9 & 4.2
    & 37.6 & 22.4 & 29.8
    & 11.4 & 5.6  & 10.8
    & 56.9 & 27.8 & 33.2
    & 51.4 & 25.9 & 31.4
    & 33.8 & 14.7 & 23.4
    & 18.9& 6.0& 18.9\\
    
    whisperv3 + Qwen 4B
    & 72.9 & 72.9 & --
    & 39.6 & 5.6 & 11.6
    & 42.9 & 26.1 & 32.7
    & 14.3 & 6.8  & 12.6
    & 61.8 & 31.4 & 37.9
    & 56.7 & 30.1 & 38.3
    & 38.6 & 17.2 & 26.6
    & 31.6& 10.7& 17.6\\
    
    whisperv3 + Qwen 8B
    & 78.1 & 78.1 & --
    & 67.4 & 10.8 & 15.4
    & 48.1 & 29.4 & 35.1
    & 17.6 & 8.2  & 14.1
    & 66.9 & 34.1 & 41.9
    & 61.8 & 34.5 & 44.2
    & 42.7 & 20.1 & 30.1
    & 43.5& 11.9& 19.5\\
    
    whisperv3 + Gemma 12B
    & 80.3 & 80.3 & --
    & 75.2 & 9.7 & 17.9
    & 50.4 & 30.8 & 36.9
    & 18.9 & 8.9  & 14.7
    & 69.1 & 35.9 & 43.1
    & 64.2 & 36.3 & 45.3
    & 44.9 & 21.6 & 31.2
    & 46.5 & 13.9 & 22.0\\
    
    whisperv3 + Gemma 27B
    & 87.9 & 87.9 & --
    & 78.9 & 10.2 & 19.5
    & 62.8 & 36.9 & 44.1
    & 26.3 & 11.9 & 18.2
    & 82.7 & 42.3 & 49.6
    & 77.1 & 44.1 & 52.3
    & 56.2 & 26.4 & 36.2
    & 50.5 & 15.7 & 25.6 \\
    
    \midrule
    
    Kimi-7B
    & 75.4 & 75.4 & --
    &  75.6& 12.4 & 15.8
    & 46.5 & 30.4 & 32.1
    & 20.5 & 9.0  & 12.3
    & 70.6 & 38.6 & 43.4
    & 65.3 & 39.1 & 45.1
    & 45.7 & 24.5 & 31.3
    & 37.8& 10.6& 16.7\\
    
    Step-Audio2-7B
    & 77.9 & 77.9 & --
    & 76.5 & 14.1 & 17.3
    & 41.4 & 26.4 & 34.1
    & 15.1 & 7.3  & 14.4
    & 64.4 & 35.4 & 43.8
    & 58.2 & 33.4 & 43.6
    & 40.8 & 18.6 & 28.2
    & 34.2& 10.4& 15.6\\
    
    Audio-Flamingo-8B
    & 64.5 & 64.5 & --
    & 75.3 & 12.4 & 15.7
    & 40.6 & 25.4 & 32.3
    & 7.9  & 4.6  & 2.3
    & 58.9 & 28.5 & 31.6
    & 44.6 & 21.6 & 31.3
    & 34.5 & 15.7 & 24.5
    & 30.4& 8.1&  14.5\\
    
    Qwen-2.5-Omni-7B
    & 79.5 & 79.5 & --
    & 80.6 & 14.7 & 19.3
    & 50.2 & 32.1 & 36.6
    & 22.4 & 10.2 & 14.5
    & 74.1 & 40.2 & 46.1
    & 68.3 & 42   & 48.2
    & 48.2 & 26.1 & 33.4
    & 36.5 & 11.2 & 17.6\\
    
    Qwen-3-Omni-30B
    & 92.4 & 92.4 & --
    & 84.6 & 15.6 & 21.7
    & 74.7 & 28.6 & 35.0
    & 33.4 & 11.3 & 23.4
    & 90.6 & 36.0 & 41.6
    & 81.5 & 24.8 & 32.4
    & 54.6 & 16.3 & 26.8
    & 41.7 & 12.4 & 19.9\\
    \bottomrule
    \end{tabular}%
    }
    \end{table*}

\begin{algorithm}[!t]
\centering
\resizebox{0.9\linewidth}{!}{
\begin{minipage}{\linewidth}
\caption{Query-conditioned audio generation with diverse speakers}
\label{alg:speaker_selection_audio_generation}
\begin{algorithmic}[1]
\Require Speaker dataset $\mathcal{D}$; embedding function $f$; subsample size $K$; queries $\mathcal{Q}$; speakers per query $m$
\Ensure Generated audio $\mathcal{A}$
\State $\mathcal{S} \leftarrow \mathrm{FPS}(\{f(s)\}_{s\in\mathcal{D}}, K)$
\State Initialize $\mathcal{A} \leftarrow \emptyset$
\ForAll{$q \in \mathcal{Q}$}
  \State Sample $m$ speakers $\mathcal{S}_q \subseteq \mathcal{S}$, $|\mathcal{S}_q|{=}m$
  \ForAll{$s \in \mathcal{S}_q$}
    \State $\tilde{a} \leftarrow \mathrm{Mix}(\mathrm{TTS}(q,s),\, n), n\sim\mathcal{N}$
    \State $\mathcal{A} \leftarrow \mathcal{A} \cup \{(q,s,\tilde{a})\}$
  \EndFor
\EndFor
\State \Return $\mathcal{A}$
\end{algorithmic}
\end{minipage}}
\end{algorithm}

The benchmark is designed to reflect realistic usage scenarios, with the taxonomy derived from tools representing real-world functionality.
To increase realism, we use state-of-the-art zero-shot voice-cloning TTS models: Qwen3TTS \cite{hu2026qwen3} and CosyVoice-3 \cite{du2025cosyvoice}. 
To capture accent diversity, we curate speakers from multiple open-source speech datasets (\Cref{tab:voice_datasets}). 
We additionally construct a background-noise set $\mathcal{N}$ comprising automotive and indoor noise recordings (e.g., engine, road, wind, HVAC, rain, turn-signal, cabin sounds, and room noise), which are mixed with the generated speech to simulate realistic deployment conditions. 
From the larger speaker pool, we select a diverse subset using stratified sampling combined with farthest point sampling (FPS), preserving accent diversity while controlling benchmark size. 
To the best of our knowledge, this is the first benchmark to provide holistic coverage of accents that affects the performance of both end-to-end and cascaded systems.
Our algorithm for voice generation is described in \Cref{alg:speaker_selection_audio_generation}.


\section{Experiments}
\label{sec:exp}
\begin{figure}
    \centering
    \includegraphics[width=1\linewidth]{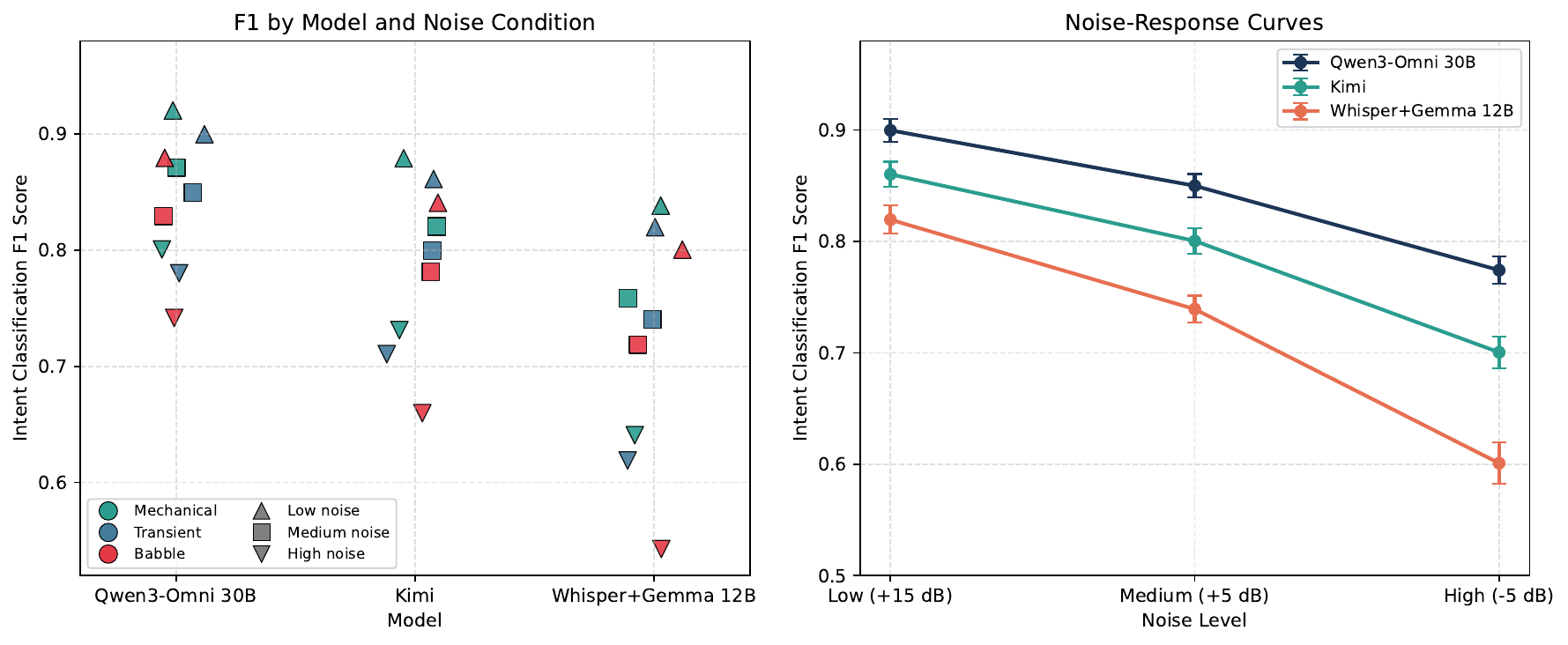}
    \caption{Effect of noise on the performance.}
    \vspace{-1em}
    \label{fig:noise_abalation}
\end{figure}
\subsection{Models}
We evaluate two categories of speech-based systems: \textit{(i) }end-to-end SpeechLMs; and \textit{(ii) }modular ASR-LLM pipelines. 
For this study, we focus exclusively on open-source models.

\noindent\textbf{SpeechLMs.}
We evaluate several state-of-the-art speech language models, including Step-Audio-2 \cite{wu2025step}, AudioFlamingo-3 \cite{goel2025audio}, Kimi Audio \cite{ding2025kimi}, Qwen-3-Omni \cite{xu2025qwen3}, and Qwen-2.5-Omni \cite{xu2025qwen3}, spanning model sizes from 7B to 30B parameters.

\noindent\textbf{ASR-LLM}
For cascaded systems, we first transcribe speech using Whisper~3 \cite{radford2023robust}, and then pass the resulting transcripts, along with tool definitions and instructions to a text-based LLM. 
We report results for multiple backbone sizes of Qwen~3 \cite{yang2025qwen3} and Gemma~3 \cite{kamath2025gemma}. 
In addition, we include an oracle setting in which Whisper transcripts are replaced with ground-truth queries to isolate the impact of ASR errors.

\subsection{Evaluation Metrics}
We evaluate zero-shot tool calling using three metrics:
\textbf{Tool Accuracy (Acc):} The fraction of examples for which the predicted tool name matches the ground truth tool. For multi-call cases, the ordered sequence of tools must match.
\textbf{Exact Match (EM):} End-to-end correctness, requiring both the predicted tool(s) and all associated arguments to exactly match the reference after deterministic normalization.
\textbf{Slot F1:} Parameter level accuracy where each argument is treated as a slot, and compute micro-averaged precision, recall, and F1 scores.

\subsection{Results}
We tabulate the results in \Cref{tab:tool_calling_results}. 
End-to-end SpeechLMs such as Qwen-3-Omni-30B and Kimi7B achieve high accuracy on simple Tier1 commands ($>$75\%), but we observe a steep decline on intermediate tiers that require multi-intent reasoning and contextual inference (i.e., Tiers 3 and 4), with EM and F1 scores often falling below 35\%. 
Tiers 5–7, which capture long-form, corrective, and multi-turn queries, remain challenging for all models, with Qwen-3-Omni-30B outperforming others but still achieving under 55\% on the F1 and EM metrics.
The most challenging scenarios involve multi-turn conversations and intent blending (Tiers 7–8), where accuracy falls below ~56\%, highlighting the difficulty of robust speech-to-tool reasoning in realistic conversational settings.
We do not evaluate Tier 8 on text-to-tool models, as intent blending mostly affects audio understanding.
Across tiers, larger models consistently outperform smaller counterparts, and tiers with multiple tools per query (i.e., Tier 3), expose weaknesses in multi-intent understanding. 
Overall, these results underscore the need for benchmarks that handle compositional tool-calling, multi-turn context, and robust argument prediction in realistic speech interactions.
Notably, end-to-end SpeechLMs do not yet consistently outperform strong ASR–LLM pipelines, indicating their effectiveness for speech-driven tool invocation.


\subsection{Ablations}
We perform ablations across three noise types - Babble Noise, Mechanical Hum and Impulsive Noise. 
We use the MS-SNSD dataset \cite{reddy2019scalable} to obtain noise files, and consider three noise levels for each type, including low (+15db SNR), medium (+5db SNR) and high noise (-5db SNR).
We notice that with addition of noise, the performance of the models deteriorate significantly as seen in Figure  \ref{fig:noise_abalation}. 
To account for this, we include different level and types of noises in our dataset, thus creating a comprehensive evaluation benchmark dataset.

\section{Conclusion}
In this paper, we introduced \textbf{Audio2Tool}, a large-scale benchmark for evaluating speech-based tool use under realistic and compositional conditions. 
Audio2Tool evaluates SpeechLMs across diverse application domains, acoustic variability, and increasing task complexity, ranging from direct commands to multi-intent and multi-turn interactions, and on many noisy environments that may require speaker disambiguation. 
Our experiments show that while current systems perform reliably on isolated commands, they struggle with compositional reasoning and accurate argument grounding.
Although we incorporate diverse synthetic voices and acoustic perturbations, the benchmark currently relies on generated speech.
We will incorporate real recordings, expand safety-critical scenarios, and broaden domain coverage in our future work. 

\section{Generative AI Use Disclosure}
During the preparation of this work, the authors used Claude Sonnet 4.5, ChatGPT, and Gemini 3, in order to generate structured figures for the overview of the pipeline shown in the paper, and for polishing the grammar of the manuscript. After using these tools, the authors reviewed and edited the content as needed and take full responsibility for the content of the publication.
\bibliographystyle{IEEEtran}
\bibliography{mybib}

\end{document}